\documentclass[twocolumn,english,pra,superscriptaddress]{revtex4-1}
\usepackage[utf8]{inputenc}
\usepackage{color}
\usepackage{babel}
\usepackage{bm}
\usepackage{graphicx}
\usepackage{physics}
\usepackage{amsmath}
\usepackage{amssymb}
\usepackage{empheq}
\usepackage{times}

\usepackage[colorlinks=true,citecolor=blue]{hyperref}
\begin{document}

\title{Quantum state tomography with tensor train cross approximation}

\author{Alexander Lidiak}
\affiliation{Department of Physics, Colorado School of Mines, Golden, Colorado 80401, USA}
\author{Casey Jameson}
\affiliation{Department of Physics, Colorado School of Mines, Golden, Colorado 80401, USA}
\author{Zhen Qin}
\affiliation{Department of Electrical \& Computer Engineering, University of Denver, Denver, Colorado 80210, USA}

\author{Gongguo Tang}
\affiliation{Department of Electrical \& Computer Engineering, University of Colorado, Boulder, Colorado 80309, USA}

\author{Michael B. Wakin}
\affiliation{Department of Electrical Engineering, Colorado School of Mines, Golden, Colorado 80401, USA}
\author{Zhihui Zhu}
\affiliation{Department of Electrical \& Computer Engineering, University of Denver, Denver, Colorado 80210, USA}
\author{Zhexuan Gong}
\email{gong@mines.edu}
\affiliation{Department of Physics, Colorado School of Mines, Golden, Colorado 80401, USA}

\begin{abstract}
It has been recently shown that a state generated by a one-dimensional noisy quantum computer is well approximated by a matrix product operator with a finite bond dimension independent of the number of qubits. We show that full quantum state tomography can be performed for such a state with a minimal number of measurement settings using a method known as tensor train cross approximation. The method works for reconstructing full rank density matrices and only requires measuring local operators, which are routinely performed in state-of-art experimental quantum platforms. Our method requires exponentially fewer state copies than the best known tomography method for unstructured states and local measurements. The fidelity of our reconstructed state can be further improved via supervised machine learning, without demanding more experimental data. Scalable tomography is achieved if the full state can be reconstructed from local reductions.
\end{abstract}

\maketitle

With the rapid development of quantum computing and quantum simulation, how to characterize and validate large quantum many-body states generated by experimental quantum devices becomes a major challenge. Among various methods \cite{aaronson2019shadow, struchalin2021experimental, smith2021efficient, huang2020predicting, harney2020entanglement}, quantum state tomography (QST) \cite{toth2010perm,cramer2010efficient,flammia2012quantum,baumgratz2013scalable,haah2017sample,Gu_2020,franca2021fast} remains the gold standard, as it provides complete information about the experimental state. However, for a generic mixed state in a $d$-dimensional Hilbert space, QST requires a number of state copies at least proportional to $d^2$ to guarantee high reconstruction fidelity \cite{haah2017sample}, thus it is generally inefficient for a quantum system of many particles. Fortunately, states generated by physical quantum systems are often structured, requiring much fewer resources in order to gain complete information about them. For example, if the quantum system is well isolated from the environment, then the state of the system is often close to a pure state and thus can be well approximated by a low-rank density matrix. In this case, one simply needs to prepare and measure the state $\sim d$ times \cite{Kai_RIP, flammia2012quantum, gross2010quantum, kyrillidis2018provable, Baldwin_CompleteMeasures_boundedrank_QST, Gu_2020, kim2021fast, gross2011recovering}. 

To make QST truly scalable, however, the number of state copies needed should only scale polynomially with the number of particles $N$. This is only possible if the state has a compact representation with only $\text{poly}(N)$ independent parameters. Examples of such states include matrix product states (MPS) \cite{schollwock2011}, matrix product operators (MPO) \cite{jarkovsky2020efficient}, tensor network states \cite{Orus2014}, and quantum neural network states \cite{torlai2018neural}. However, having an efficient representation of the quantum state does not imply that an efficient QST method exists. For a pure state represented by an MPS with a finite bond dimension, efficient QST methods have been found and tested for various physical states \cite{cramer2010efficient,wang2020scalable,lanyon2017efficient}, although a rigorous bound on the number of state copies needed to guarantee high fidelity QST is yet to be found. States that can be efficiently represented by artificial neural networks have also been targeted for QST \cite{carrasquilla2019reconstructing, schmale2021scalable, rocchetto2018learning, Torlai_exp_QST, xin2019local, palmieri2020experimental, rocchetto2018learning, torlai2018neural, Torlai_NDO, ahmed2021quantum, lange2022adaptive, quek2021adaptive}, but it is not clear whether efficient QST schemes exist in general for such states.

In this paper, we focus on QST for a mixed state represented by an MPO with a finite bond dimension independent of the system size. Such an MPO has recently been shown to describe most states generated by a one-dimensional quantum computer with a finite error rate for elementary quantum gates \cite{noh2020efficient}. Given the popularity of one-dimensional quantum computers such as those based on trapped ions \cite{MonroeRMP}, efficient QST methods for states generated by such devices are highly desired. However, no guaranteed efficient method to perform QST with bounded error on these states has been developed. We also note that such states can have very high entropy, making QST methods designed for low-rank density matrices generally not applicable. The method currently demonstrating highest efficiency is developed in Ref.\,\cite{baumgratz2013scalable}, which attempts to reconstruct the MPO representing the target state from local reduced density matrices each involving at most $R$ particles. The method requires performing full tomography on the reduced density matrices, which involves a number of state copies exponential in $R$. If $R$ is small, this method is efficient, but $R$ cannot be determined without knowing the state explicitly or even predicted from the bond dimension, and thus the efficiency of the whole method is not guaranteed.

Here we take an important step towards addressing this problem. Using a method in signal processing and compressed sensing know as tensor train cross approximation \cite{oseledets2010tt,savostyanov2011fast,savostyanov2014quasioptimality,Zhen_Zhihui_Error}, we show that an $N$-particle state represented by an MPO with a finite constant bond dimension can be reconstructed by measuring the state in only $O(N)$ different bases. To our best knowledge, the application of cross approximation to QST has not been studied in depth before \cite{Petz_2018}. Importantly, cross approximation only requires local measurements on individual quantum particles, which are routinely performed in current quantum experiments. For example, on qubit systems, we need only measure one of the three Pauli operators $\sigma^{x,y,z}$ for each qubit. As a result, our method is easily implementable experimentally. For a generic mixed state made of $N$ qubits, one needs to measure all $3^N$ different combinations of local Pauli operators in order to gain complete information of the state \cite{Gu_2020}. Therefore, our method requires only a small fraction $O(N)/3^N$ of measurement bases (as well as the number of state copies) compared to such unstructured tomography. We emphasize that our $O(N)$ scaling of the number of measurement bases is optimal since the target MPO state contains $O(N)$ independent parameters.

Nevertheless, our method does not guarantee that the number of total state copies is polynomial in $N$. Due to the statistical errors in quantum measurements, an exponentially large number of copies of states per measurement basis needs to be used to ensure a bounded error (in the Hilbert-Schmidt norm) of the reconstructed state. Without additional assumptions on the target MPO state, we expect this limitation to be fundamental, as it also applies to the best-known unstructured tomography method using local measurements \cite{haah2017sample}. We develop a supervised machine learning method to alleviate the effects of statistical errors and the requirement of a large number of measurements. Finally, we point out that we can combine our method with the protocol in Ref.\,\cite{baumgratz2013scalable} to achieve efficient QST if a local reduction exists for the target MPO state. Our method can perform QST on the reduced states much more efficiently than standard methods, thus improving the protocol in Ref.\,\cite{baumgratz2013scalable} without affecting its scalability condition.

This paper is organized as follows: In Section I, we introduce the method of cross approximation for both matrices and tensors. Section II applies cross approximation to QST for physical target states represented by MPOs. In Section III, we analyze the effects of statistical errors in quantum measurements. We then show that the detrimental effects of such measurement errors can be further reduced via a supervised machine learning method in Section IV. The paper ends with a discussion and outlook section.

\section{Introduction of Cross approximation}
The tensor train cross approximation we use for QST is a generalization of the cross approximation for a matrix, which is also known as \emph{skeleton decomposition} or \emph{CUR decomposition} \cite{boutsidis2017optimal,mitrovic2013cur,aldroubi2019cur,xu2015cur}. The main idea of matrix cross approximation is that given a low rank matrix, we can possibly approximate it using a small number of its rows and columns. Formally, following the standard notation for the CUR decomposition, we express the cross approximation for a general $m\times n$ complex valued matrix $\mathbf{A}\in\mathbb{C}^{m\times n}$ as
\begin{align}
    \label{eq:CA_matrix}
    \mathbf{A} \approx \mathbf{C} \mathbf{U}^+ \mathbf{R},
\end{align}
where $\mathbf{C}=\mathbf{A}(:,J)$, $\mathbf{R}=\mathbf{A}(I,:)$, and $\mathbf{U}=\mathbf{A}(I,J)$ with  $I$ and $J$ respectively denoting some subsets of the indices of $\mathbf{A}$'s rows and columns. $\mathbf{U}^{+}$ denotes the pseudo inverse of the matrix $\mathbf{U}$ (not to be confused with Hermitian conjugate).

\begin{figure}[ht]
   \centering
   \includegraphics[width =\columnwidth]{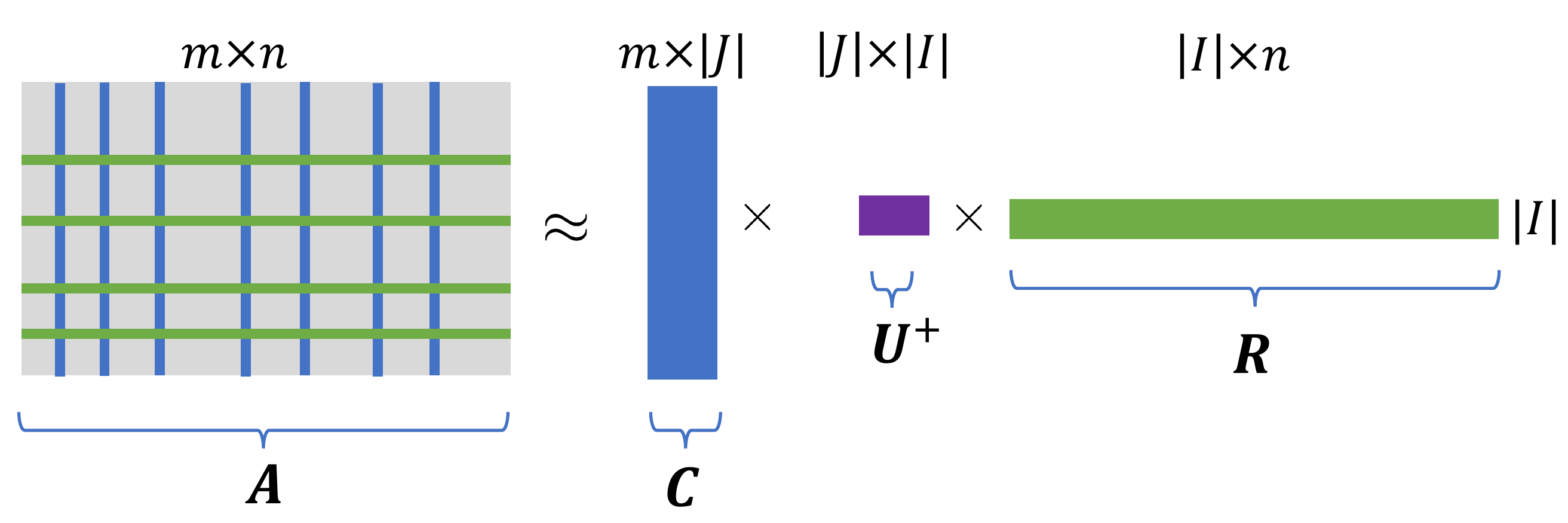}
   \caption{Illustration of matrix cross approximation. An $m\times n$ matrix $\mathbf{A}$ is approximated by a product of three matrices $\mathbf{C}$, $\mathbf{U}^+$, and $\mathbf{R}$, where $\mathbf{C}$ is formed by selecting $|J|$ columns from $\mathbf{A}$, $\mathbf{R}$ is formed by selecting $|I|$ rows from $\mathbf{A}$, and $\mathbf{U}$ is formed by the union of $\mathbf{C}$ and $\mathbf{R}$.}
   \label{fig:ca_matrix}
\end{figure}

An illustration of the matrix cross approximation is shown in Fig.\,\ref{fig:ca_matrix}. One can mathematically prove that if $\text{rank}(\mathbf{U})=\text{rank}(\mathbf{A})$, then the cross approximation becomes exact, i.e. $\mathbf{A} = \mathbf{C} \mathbf{U}^{-1} \mathbf{R}$. In this case, one can also perform a singular value decomposition (SVD) of $\mathbf{A}$ to obtain an exact decomposition of $\mathbf{A}$ in a similar form. However, computing the SVD requires  full knowledge of the matrix $\mathbf{A}$, while the cross approximation only requires a small fraction of $\mathbf{A}$'s rows and columns be known for a low rank matrix. As we will show, this advantage of cross approximation allows us to measure only a small number of observables for a target quantum state represented by a compact MPO.

If the rank of the matrix $\mathbf{A}$ is not known a priori, then cross approximation is in general less accurate and robust than the SVD, which provides the best approximation of $\mathbf{A}$ for a chosen rank. For cross approximation to be optimal, one needs to choose the rows and columns of $\mathbf{A}$ in a way that maximizes the volume (i.e. determinant in modulus) of $\mathbf{U}$ in Eq.\,\eqref{eq:CA_matrix}. The search for the maximum volume submatrix $\mathbf{U}$ is in general computationally expensive, but many efficient methods have been developed to achieve quasi-optimal results \cite{civril2009selecting,deshpande2010efficient,deshpande2006matrix,cortinovis2020low,goreinov2001maximal,chen2015completing}.

We now generalize the matrix cross approximation to tensor train cross approximation. We denote an order-$N$ tensor $\mathcal{A}$'s elements using $\mathcal{A}(\gamma_1,\gamma_2,\cdots,\gamma_N$), where $\gamma_i=0,1,2,\cdots d_i-1$ is the index for the $i^{\text{th}}$ dimension of the tensor. We focus on the case where $\mathcal{A}$ has a tensor train decomposition of the form:
\begin{align}
\mathcal{A}(\gamma_{1},\cdots\gamma_{N})= G_{1}^{\gamma_{1}}G_{2}^{\gamma_{2}}\cdots G_{N}^{\gamma_{N}}
\label{eq:CA_tt}
\end{align}
where each $G_i^{\gamma_i}$ is a matrix of dimension $\chi_{i-1} \times \chi_i$ with $\chi_0=\chi_N=1$. We point out that the tensor train in Eq.\,\eqref{eq:CA_tt} is identical to an MPS or MPO when the tensor $\mathcal{A}$ represents a wavefunction or density matrix of a quantum many-spin system \cite{schollwock2011}, with $\{\chi_i\}$ often referred to as the \emph{bond dimensions}.

Given an arbitrary tensor $\mathcal{A}$, one can always find the matrices $\{G_i^{\gamma_i}\}$ in the tensor train decomposition Eq.\,\eqref{eq:CA_tt} via successive SVDs \cite{schollwock2011}. Such successive SVDs further allow one to compress the dimension $\{\chi_i\}$ of the the matrices $\{G_i^{\gamma_i}\}$ \cite{schollwock2011}. However, the entire tensor $\mathcal{A}$, which contains exponentially many elements in $N$, needs to be known in order to perform the SVD, which is impractical for large $N$. 

\begin{figure}[ht]
   \centering
   \includegraphics[width =\columnwidth]{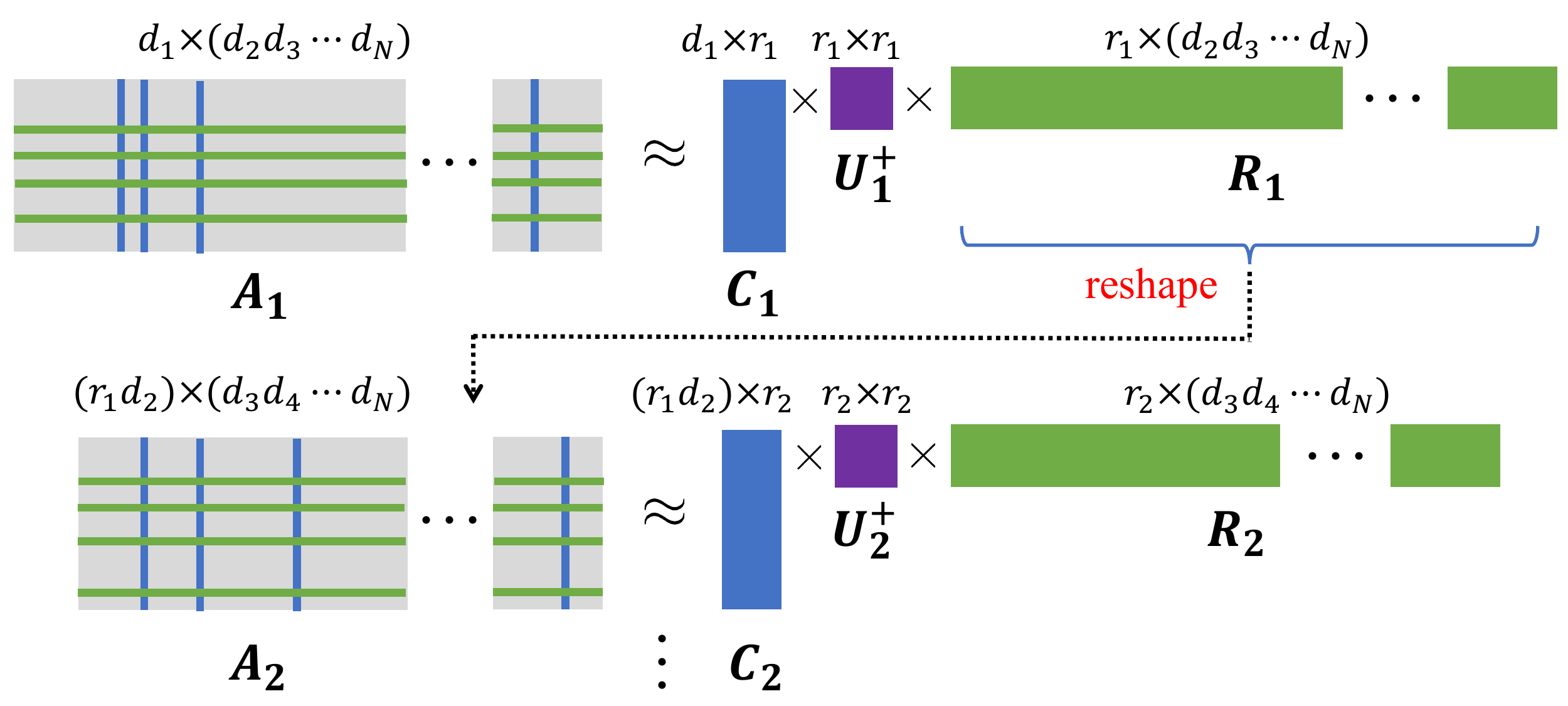}
   \caption{Illustration of tensor train cross approximation. In step one, we perform a matrix cross approximation of $\mathbf{A}_1$, which is obtained from reshaping the tensor $\mathcal{A}$. In step two, we reshape the matrix $\mathbf{R}_1$ obtained in step 1 into a matrix $\mathbf{A}_2$ and perform a matrix cross approximation of $\mathbf{A}_2$. We iterate this step for a total of $N-1$ steps.}
   \label{fig:ca_tensor}
\end{figure}

Similar to the matrix case, we can apply cross approximation instead of SVD for the tensor $\mathcal{A}$ to improve the efficiency of the tensor train decomposition. This consists of the following steps (illustrated in Fig.\,\ref{fig:ca_tensor}). In the first step, we reshape the tensor $\mathcal{A}$ with dimensions $d_1\times d_2 \times \cdots \times d_N$ into a wide matrix $\mathbf{A}_1$ of dimension $d_1 \times (d_2 d_3 \cdots d_N)$ and perform cross approximation on this matrix by selecting $r_1$ rows and columns of $\mathbf{A}_1$, resulting in $\mathbf{A}_1 \approx \mathbf{C}_1 \mathbf{U}_1^+ \mathbf{R}_1$. One can then show that $\mathbf{A}_1$ has a rank of at most $\chi_1$, and thus if $r_1\ge \chi_1$, the matrix cross approximation of $\mathbf{A}_1$ can be exact. The matrix $\mathbf{C}_1 \mathbf{U}_1^+$, which has a dimension of $d_1 \times r_1$, creates $d_1$ row vectors $\{\tilde{G}_1^{\gamma_1}\}$ each of dimension $r_1$.

In the second step, we reshape the previously obtained matrix $\mathbf{R}_1$ into a matrix $\mathbf{A}_2$ of dimension $(r_1 d_2) \times (d_3 d_4 \cdots d_N)$ and perform a cross approximation of $\mathbf{A}_2 \approx \mathbf{C}_2 \mathbf{U}_2^+ \mathbf{R}_2$ by choosing $r_2$ rows and columns. Again, one can show that the matrix $\mathbf{A}_2$ has a rank of at most $\chi_2$, thus the cross approximation of $\mathbf{A}_2$ can be exact if $r_2\ge \chi_2$. The matrix $\mathbf{C}_2 \mathbf{U}_2^+$, which has a dimension of $(r_1 d_2) \times r_2$, can be reshaped to create $d_2$ matrices $\{\tilde{G}_2^{\gamma_2}\}$ each of dimension $r_1\times r_2$.

We iterate the above step for a total of $N-1$ steps. This allows us to obtain a set of matrices $\{\tilde{G}_i^{\gamma_i}\}$ for $i=1,2,\cdots N$ and $\gamma_i = 1,2,\cdots d_i-1$, where in the last step the matrix $\mathbf{R}_{N-1}$ is used to obtain $\{\tilde{G}_N^{\gamma_N}\}$. As we mentioned, if we choose $r_i \ge \chi_i$ in the matrix cross approximation in step $i$, we can exactly recover the original tensor as $\mathcal{A}(\gamma_{1},\cdots\gamma_{N})= \tilde{G}_{1}^{\gamma_{1}}\tilde{G}_{2}^{\gamma_{2}}\cdots \tilde{G}_{N}^{\gamma_{N}}$. Most importantly, this process is highly efficient, as in each step we only need to find the small matrices $\mathbf{C}_i$ of dimension $(r_{i-1} d_{i}) \times r_i$ and $\mathbf{U}_i$ of dimension $r_i \times r_i$, and they can be constructed directly from the elements of the tensor $\mathcal{A}$. Assuming that $d=\max_i d_i$ and $r=\max_i r_i$, then we only need to know approximately $N d r^2$ elements of $\mathcal{A}$ to fully reconstruct it. And in the case of $r_i=\chi_i$, this reconstruction method is optimal because $\mathcal{A}$ contains the same number of independent parameters.

In reality, $\mathcal{A}$ may not have an exact tensor train decomposition as in Eq.\,\eqref{eq:CA_tt}, or we may not know the bond dimension $\{\chi_i\}$ of the exact tensor train decomposition. In this case, we can choose $r_i$ in the tensor train cross approximation heuristically, starting from a small initial guess. If $r_i < \chi_i$, then the cross approximation is no longer exact but we can use maximum volume principle based algorithms, of which two have been developed, to find the optimal entries of the tensor $\mathcal{A}$ used to perform the tensor train decomposition. The first algorithm is introduced in Ref.\,\cite{savostyanov2011fast} and known as the ``DMRG-cross" algorithm, since it is similar to the density matrix renormalization group (DMRG) method used in variationally finding the ground state of a 1D quantum many-body system \cite{schollwock2011}. Similar to DMRG, it can choose $r_i$ adaptively between 1 and a preset maximum value based on a chosen local error threshold. The second is introduced in Ref.\,\cite{savostyanov2014quasioptimality} and is known as the ``greedy-cross" algorithm, which uses the method of greedy approximation \cite{temlyakov_2008} and chooses an aggressively small $r_i$. We will mainly use the DMRG-cross algorithm for QST as it performs more reliably in our calculations.

\section{Cross approximation based QST}
We now apply the above-mentioned tensor train cross approximation method to quantum state tomography. For simplicity, we focus on a quantum state of $N$ qubits, but it is straightforward to generalize our method to a quantum system of $N$ qudits. A general mixed state of $N$ qubits can be described by a density operator $\rho$ in the $N$-qubit Pauli operator basis.
\begin{equation}
\rho= \sum_{\gamma_1,\cdots,\gamma_N}\mathcal{A}(\gamma_1,\gamma_2,\cdots \gamma_N)\left(\sigma_{1}^{\gamma_{1}}\sigma_{2}^{\gamma_{2}}\cdots\sigma_{N}^{\gamma_{N}}\right)
\label{eq:rho}
\end{equation}
where each $\gamma_{i}=0,1,2,3$ and $\sigma_i^{0,1,2,3}$ denotes respectively the identity operator and Pauli operator $\sigma_i^{x,y,z}$ acting non-trivially only on qubit $i$. As a result, the full state is described by the order-$N$ tensor $\mathcal{A}$ we discussed in Section I with $d=d_1=d_2\cdots=d_N=4$. Importantly, each element of $\mathcal{A}$ can be measured experimentally since
\begin{equation}
\mathcal{A}(\gamma_{1},\cdots\gamma_{N})=\langle \sigma_{1}^{\gamma_{1}}\sigma_{2}^{\gamma_{2}}\cdots\sigma_{N}^{\gamma_{N}}\rangle/2^{N}.
\label{eq:A}
\end{equation}

For state-of-art quantum computers and quantum simulators, the expectation value of $\sigma_{1}^{\gamma_{1}}\sigma_{2}^{\gamma_{2}}\cdots\sigma_{N}^{\gamma_{N}}$ can be measured by locally measuring each qubit (spin-1/2) in the $X$, $Y$, or $Z$ direction in an arbitrary order. Note that if $\gamma_i=0$, one can measure the qubit $i$ in any direction (or not measure it) and simply replace the operator $\sigma_i^0$ by unity in evaluating the expectation value.

If $\mathcal{A}$ for our target state can be represented or well approximated by a tensor train (or MPO) defined in Eq.\,\eqref{eq:CA_tt} with a maximum bond dimension $\chi\equiv\max_i \chi_i$, then we can perform QST by reconstructing $\mathcal{A}$ (and hence the full state) using tensor-train cross approximation. The main advantage of this QST method is that only about $4N \chi^2$ elements of $\mathcal{A}$ need to be measured, which is also the number of different measurement bases required. As we mentioned in Section I, this is the minimal number of measurement bases required to gain full information of the target state.

Another advantage of our QST protocol is that we obtain an efficient representation of the target state in an MPO form (consisting of the matrices $\{\tilde{G}_i^{\gamma_i}\}$). The full density matrix of the target state is never reconstructed or stored, explicitly. However, the reconstructed MPO can be used to compute the expectation values of most physically interesting observables efficiently \cite{Verstraete2004}. In particular, these observables include $N$-body correlation functions, such as $\sigma_{1}^{\gamma_{1}}\sigma_{2}^{\gamma_{2}}\cdots\sigma_{N}^{\gamma_{N}}$ with no $\gamma_i=0$. The expectation value of such global observable is hard to obtain using shadow tomography techniques \cite{aaronson2019shadow,huang2020predicting} and full QST is usually needed.

We now benchmark the performance of cross approximation based QST using two different classes of physical target states represented by MPOs with a given bond dimension. First, we consider thermal states of a 1D quantum Ising model, described by the Hamiltonian
\begin{equation}
H=\sum_{i=1}^{N-1}\sigma_{i}^{z}\sigma_{i+1}^{z}+g\sum_{i=1}^{N}\sigma_{i}^{x}.
\label{eq:HTFIM}
\end{equation}
We will set $g=1$, which makes the ground state of $H$ at its quantum critical point in the thermodynamic limit. Such ground state requires the largest bond dimension among all values of $g$ for it to be approximated by an MPS or MPO. The thermal state of $H$ is defined by $\rho_{T}=\frac{e^{-H/T}}{\text{Tr}(e^{-H/T})}$, and it can be well approximated by an MPO with a small bond dimension for tens or even hundreds of spins \cite{Verstraete2004}. We will set $T=0.2$ (corresponding to a low temperature state close to the ground state) and $T=2$ (corresponding to a high temperature state) in our following calculations. We generate $\rho_{T}$ in an MPO form for up to $N=40$ qubits using the Open Source Matrix Product State (OSMPS) software package \cite{OSMPS,OSMPS_Jaschke}, which uses an imaginary time evolution of an MPO ansatz to approximate $\rho_{T}$ with a maximum bond dimension of $32$.

Our second class of target states are random locally purified tensor network (LPTN) states \cite{Werner2016}. These states are represented by random MPOs that are guaranteed to be physical. The density operator $\rho_{\text{LPTN}}$ of such a random LPTN state is represented by
\begin{equation}
\langle s_{1}\cdots s_{N}| \rho_{\text{LPTN}} | s_{1}^{\prime}\cdots s_{N}^{\prime} \rangle = \mathbf{M}_{1}^{s_{1},s_{1}^{\prime}}\cdots \mathbf{M}_{N}^{s_{N},s_{N}^{\prime}}
\label{eq:LPTN}
\end{equation}
where $|s_1 \cdots s_N\rangle$ with $s_i=0,1$ denotes the computational basis state for $N$ qubits. To make sure that the above density operator $\rho$ is physical, it must be semi-positive definite. This can be guaranteed if each matrix $\mathbf{M}_{i}^{s_{i},s_{i}^{\prime}}$ takes the following form \cite{Verstraete2004}:
\begin{equation}
\mathbf{M}_{i}^{s_{i},s_{i}^{\prime}}=\sum_{a_{i}=1}^{K_i}\mathbf{A}_{i}^{s_{i},a_{i}}\otimes(\mathbf{A}_{i}^{s_{i}^{\prime},a_{i}})^{\ast}
\end{equation}
where $^{\ast}$ denotes complex conjugate and $K_i$ is an arbitrary positive integer. For simplicity, we assume each matrix $\mathbf{A}_{i}^{s_{i},a_{i}}$ is of dimension $\kappa\times\kappa$,
except that $\mathbf{A}_{1}^{s_{1},a_{1}}$ and $\mathbf{A}_{N}^{s_{N},a_{N}}$ are of
dimension $1\times\kappa$ and $\kappa\times1$ respectively. To make the state $\rho_{\text{LPTN}}$ sufficiently random, we set each matrix element of $\mathbf{A}_{i}^{s_{i},a_{i}}$
to be a random complex number with both its real and imaginary parts drawn uniformly from $[-1,1]$. In addition, we set $K_i=10$ to make sure the state is sufficiently mixed ($K_i=1$ will generate a pure state).

It is straightforward to see that $\rho_{\text{LPTN}}$ is an MPO with bond dimension $\chi=\kappa^{2}$ 
and is always Hermitian. However, to make $\rho_{\text{LPTN}}$ physical, it still needs to be normalized. This is done by calculating
\begin{equation}
\text{Tr}(\rho)=(\mathbf{M}_{1}^{0,0}+\mathbf{M}_{1}^{1,1})\cdots(\mathbf{M}_{N}^{0,0}+\mathbf{M}_{N}^{1,1})
\end{equation}
and dividing each matrix $\mathbf{M}_{i}^{s_{i},s_{i}^{\prime}}$ matrix above by $[\text{Tr}(\rho)]^{1/N}$.

Finally, we need to convert the MPO $\{\mathbf{M}_{i}^{s_{i},s_{i}^{\prime}}\}$ representing $\rho_{\text{LPTN}}$ from the computational basis to the Pauli operator basis to comply with our measurement scheme. This can be done using the following linear transformation that preserves the bond dimension of the MPO:
\begin{align}
G_{i}^{0} & =\frac{M_{i}^{0,0}+M_{i}^{1,1}}{2}\qquad G_{i}^{1}=\frac{M_{i}^{0,1}+M_{i}^{1,0}}{2}\\
G_{i}^{2} & =i\frac{M_{i}^{0,1}-M_{i}^{1,0}}{2}\qquad G_{i}^{3}=\frac{M_{i}^{0,0}-M_{i}^{1,1}}{2}.
\end{align}

To quantify the performance of tensor train cross approximation, we also need to define a distance measure between the reconstructed state and the original target state. Commonly used distance measures for quantum states include trace distance and fidelity \cite{nielsen_quantum_2010}, but both of these measures cannot be computed efficiently for a large number of qubits even if the states have a compact MPO representation. Here we instead use a normalized Frobenius norm (squared) difference as a distance measure between two density matrices $\rho_1$ and $\rho_2$, defined as \cite{baumgratz2013scalable}
\begin{equation}
D(\rho_{1},\rho_{2})\equiv \frac{\lVert\rho_{\text{1}}-\rho_{2}\rVert_{2}^{2}}{\lVert\rho_{1}\rVert_{2}^{2}}=\frac{\text{Tr}(\rho_{1}^{\dagger}\rho_{1}+\rho_{2}^{\dagger}\rho_{2}-\rho_{1}^{\dagger}\rho_{2}-\rho_{2}^{\dagger}\rho_{1})}{\text{Tr}(\rho_{1}^{\dagger}\rho_{1})}
\label{eq:D}
\end{equation}
where $\dagger$ denotes Hermitian conjugate and $\lVert A\rVert_{2}=\sqrt{\text{Tr}(A^{\dagger}A)}$ denotes the Frobenius norm of any matrix $A$. Importantly, $D(\rho_{1},\rho_{2})$ can be computed efficiently if both $\rho_1$ and $\rho_2$ are represented by compact MPOs. For example, if
\begin{align}
\rho_{1} & =\sum_{\gamma_{1},\cdots,\gamma_{N}}\left(G_{1}^{\gamma_{1}}G_{2}^{\gamma_{2}}\cdots G_{N}^{\gamma_{N}}\right)\sigma_{1}^{\gamma_{1}}\sigma_{2}^{\gamma_{2}}\cdots\sigma_{N}^{\gamma_{N}}\label{eq:rho1}\\
\rho_{2} & =\sum_{\gamma_{1},\cdots,\gamma_{N}}\left(\tilde{G}_{1}^{\gamma_{1}}\tilde{G}_{2}^{\gamma_{2}}\cdots\tilde{G}_{N}^{\gamma_{N}}\right)\sigma_{1}^{\gamma_{1}}\sigma_{2}^{\gamma_{2}}\cdots\sigma_{N}^{\gamma_{N}},\label{eq:rho2}
\end{align}
then terms such as $\text{Tr}(\rho_{1}^{\dagger}\rho_{2})$ in Eq.\,\eqref{eq:D} can be computed efficiently using contractions of the two MPOs:
\begin{align}
\frac{\text{Tr}(\rho_{1}^{\dagger}\rho_{2})}{2^N} & =\left(\sum_{\gamma_{1}}(G_{1}^{\gamma_{1}})^{\ast}\otimes\tilde{G}_{1}^{\gamma_{1}}\right)\cdots\left(\sum_{\gamma_{N}}(G_{N}^{\gamma_{N}})^{\ast}\otimes\tilde{G}_{N}^{\gamma_{N}}\right).
\end{align}

The distance measure $D$ can only be calculated for a numerical benchmark experiment. For an actual quantum experiment, we do not have full knowledge of the state (or its MPO representation) and therefore cannot calculate $D$. However, we can guess if our QST protocol succeeded in well approximating the experimental target using a sampled version of $D$, denoted by $D_s$ below, to quantify the quality of the QST:
\begin{equation}
D_s(\rho_{1},\rho_{2})=\frac{\sum^{\prime}_{\gamma_{1}\cdots\gamma_{N}}\left|\langle\sigma_{1}^{\gamma_{1}}\cdots\sigma_{N}^{\gamma_{N}}\rangle_{\rho_{1}}-\langle\sigma_{1}^{\gamma_{1}}\cdots\sigma_{N}^{\gamma_{N}}\rangle_{\rho_{2}}\right|^{2}}{\sum^{\prime}_{\gamma_{1}\cdots\gamma_{N}}\left|\langle\sigma_{1}^{\gamma_{1}}\cdots\sigma_{N}^{\gamma_{N}}\rangle_{\rho_{1}}\right|^{2}}
\label{eq:Ds}
\end{equation}
where $\sum^{\prime}_{\gamma_{1}\cdots\gamma_{N}}$ denotes the sum over only the indices $\{\gamma_1,\cdots\gamma_N\}$ used by the cross approximation (and the corresponding local bases in which measurements are performed). Since we only need $O(N)$ measurement bases, $D_s$ can be computed easily from experimental measurement data.

The difference between the target state and the cross approximation reconstructed state comes primarily from two sources: (1)~Underestimation of the bond dimension of the target MPO state leads the cross approximation to be inexact. (2)~Statistical error in quantum measurements leads the elements of the tensor $\mathcal{A}$ (and therefore the inputs to the cross approximation) to be inexact. For the rest of this section, we will focus on the first error source by assuming zero statistical error in the measurement. This is of course an impractical assumption as it will require an infinite number of state copies per measurement basis. We will remove this assumption and focus on the effects of the statistic error in Section III.

We have performed numerical experiments for synthetic target states, those being the aforementioned thermal states of Eq.\,\eqref{eq:HTFIM} and the random LPTN states in Eq.\,\eqref{eq:LPTN}. For both types of target states, we can efficiently calculate the expectation values of $\sigma_{1}^{\gamma_{1}}\sigma_{2}^{\gamma_{2}}\cdots\sigma_{N}^{\gamma_{N}}$  \cite{Verstraete2004} needed for the tensor train cross approximation. We then apply the DMRG-cross algorithm to reconstruct the target state for up to $N=40$ qubits. Without fine tuning, we set the maximum bond dimension used in DMRG-cross to $10$, and the local truncation error to $10^{-3}$.

\begin{figure}[ht]
   \centering
   \includegraphics[width=0.49\columnwidth]{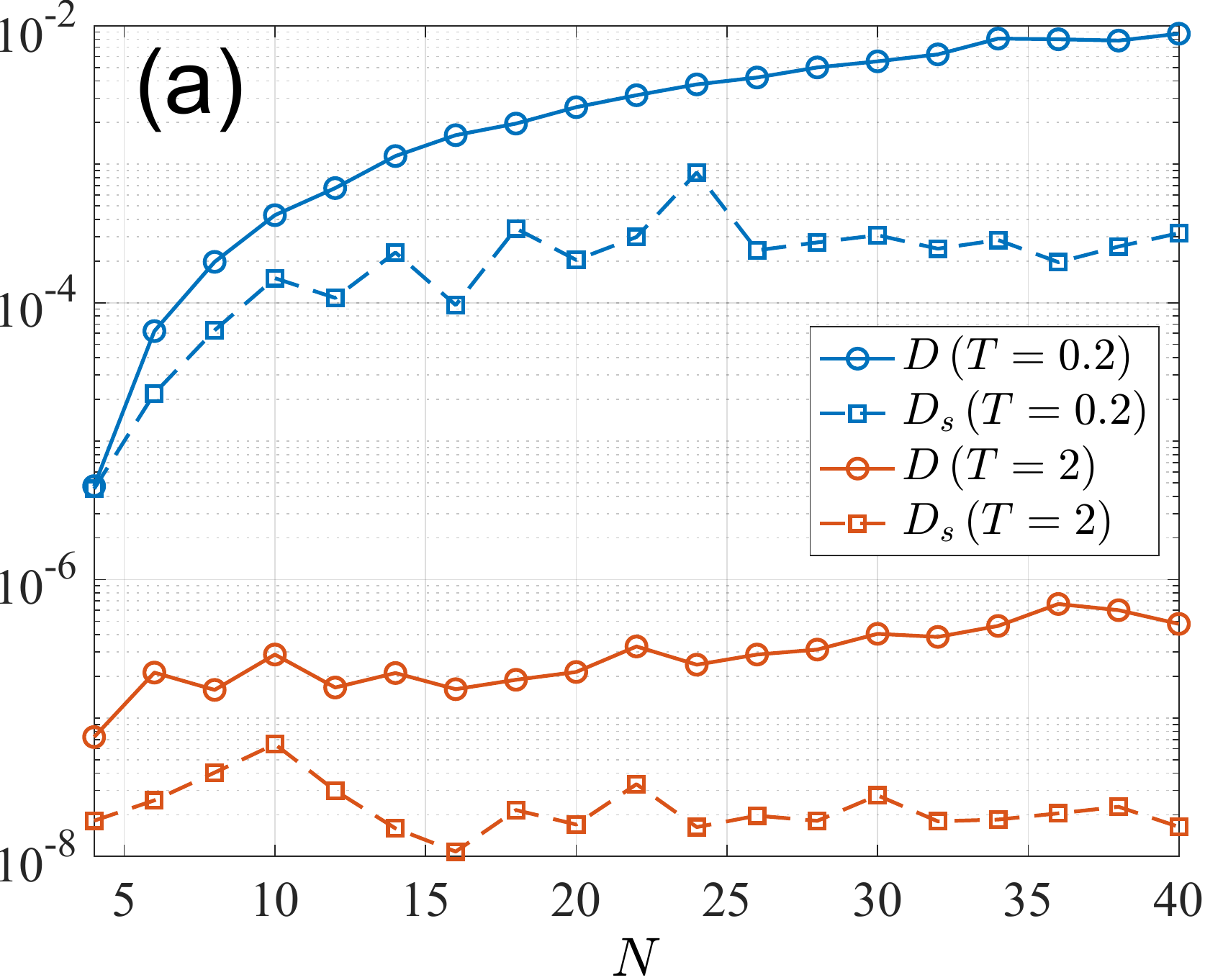} \hfill
   \includegraphics[width=0.49\columnwidth]{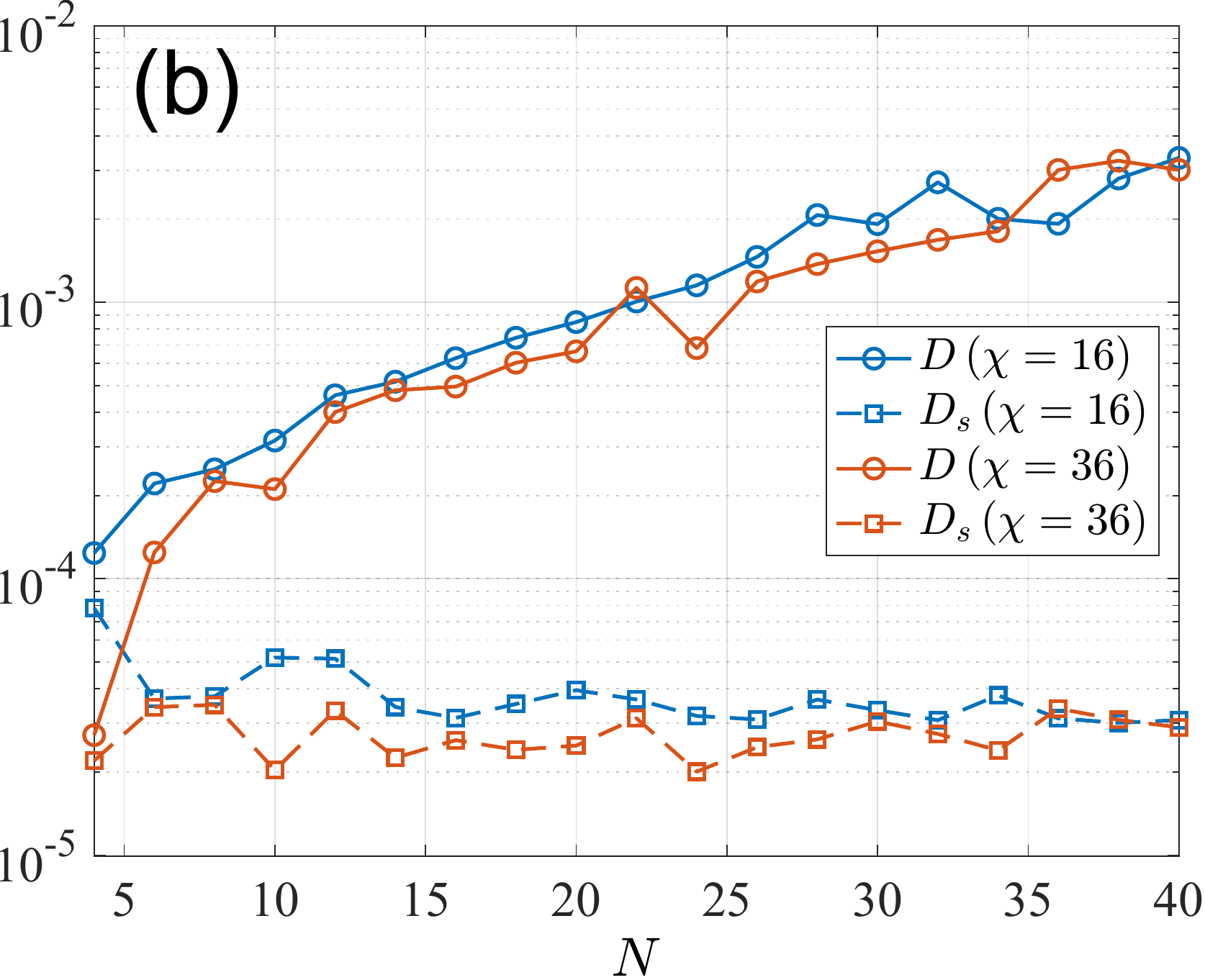}

   \caption{Distance measures ($D$ and $D_s$) between the target states and the reconstructed states using tensor train cross approximation without measurement errors as a function of the number of qubits $N$. The target states are thermal states of a quantum Ising model with $T=0.2$ or $2$ (a) and random LPTN states with $\chi=16$ or $36$ (b).}
   \label{fig:ca_D}
\end{figure}

As expected, we find that the cross approximation works very well (with $D<10^{-6}$) when the DMRG-cross algorithm does not underestimate the bond dimension of the target state. This is the case for the thermal states of Eq.\,\eqref{eq:HTFIM} with $T=2$ (see Fig.\,\ref{fig:ca_D}a), which are well approximated by MPOs with bond dimensions not exceeding 7 for $N\le40$, and DMRG-cross adaptively chooses bond dimensions between 6 and 8 for these states. Ideally, we expect $D$ to be zero (or at machine error level) if we do not underestimate the bond dimension, but numerical instabilities associated with the pseudo-inverse could account for the nonzero $D$ observed \cite{Zhen_Zhihui_Error}.

For most other target states, DMRG-cross underestimates the bond dimensions. But remarkably, the cross approximation still works well in such scenarios, achieving $D<10^{-2}$ everywhere in Fig.\,\ref{fig:ca_D}. This means that the tensor train cross approximation can actually be used as a compression technique for MPOs. Unlike most other MPO compression techniques, the cross approximation does not have any restrictions on the locality of MPOs \cite{jarkovsky2020efficient}, and is particularly useful for compressing quantum states represented by MPOs. Specifically, the thermal states with $T=0.2$ are represented by MPOs of increasing bond dimension in $N$, up to $32$. And we clearly see that $D$ increases with $N$ due to an increasingly underestimated bond dimension. This is also the case for the random LPTN states, whose bond dimensions are either $16$ or $36$. Naively, one would expect the errors due to the underestimated bond dimensions would proliferate exponentially in $N$ due to the iterative process in tensor train cross approximation (see Fig.\,\ref{fig:ca_tensor}). However, we only see a polynomial increase of the distance measure $D$ as $N$ increases. This observation is consistent with recent mathematical results on the error bounds of tensor train cross approximation showing that the reconstruction error in general grows only polynomially in $N$ \cite{savostyanov2014quasioptimality,Zhen_Zhihui_Error}. For all target states we studied, the sampled distance $D_s$ is generally well below the value of $D$.

\begin{figure}[ht]
   \centering
   \includegraphics[width=0.49\columnwidth]{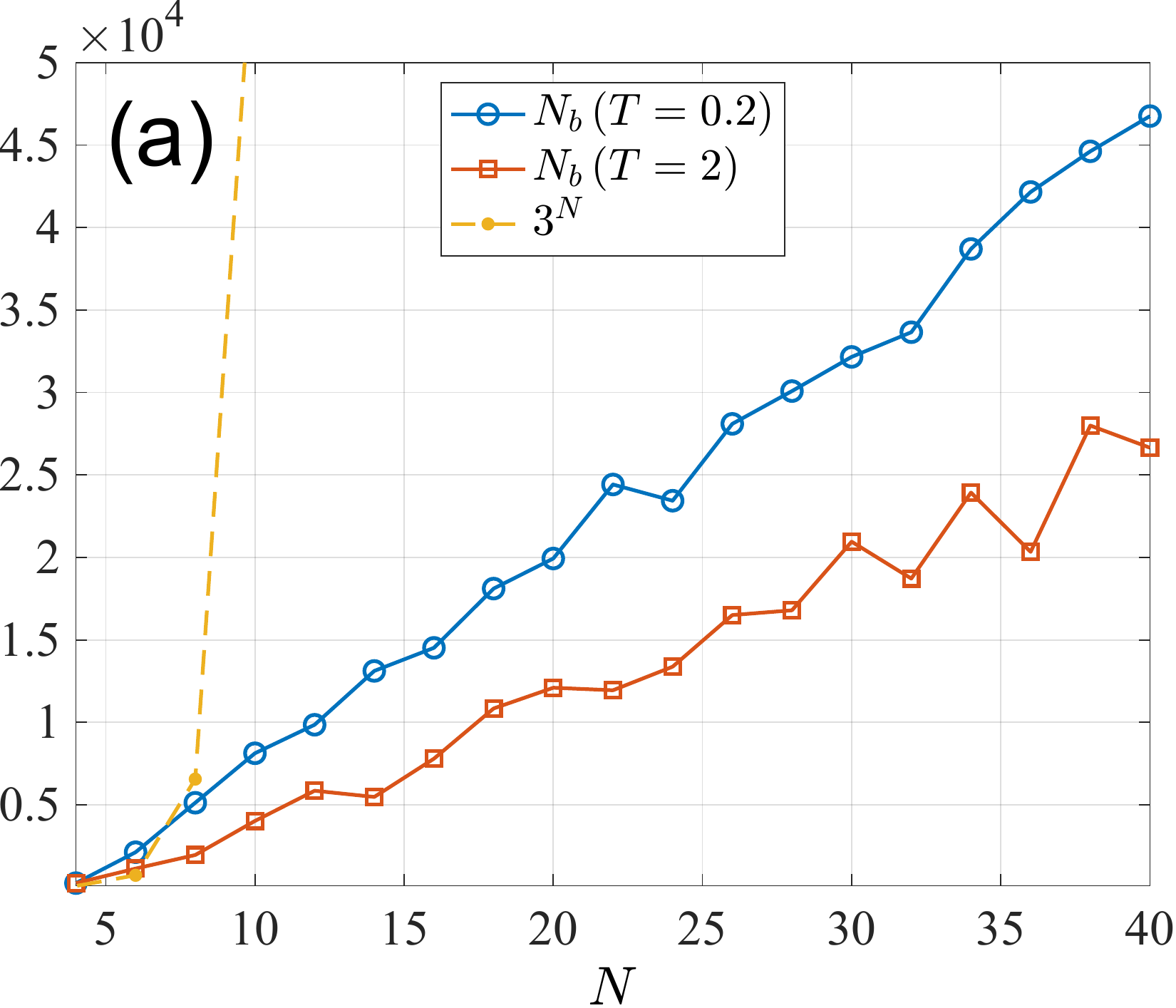} \hfill
   \includegraphics[width=0.49\columnwidth]{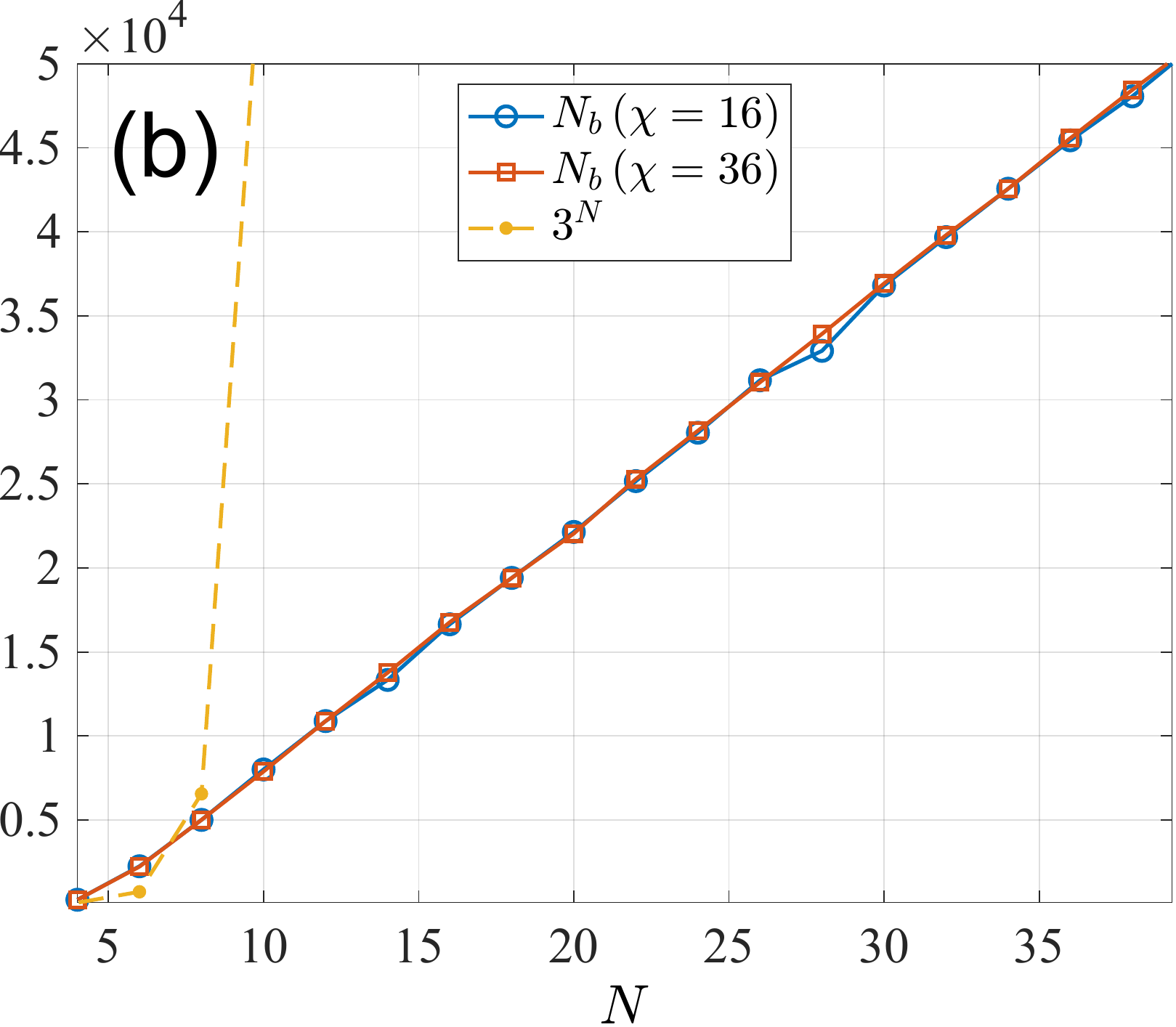}
   \caption{The number of measurement bases $N_b$ required by the tensor train cross approximation as a function of $N$. We also plot the number of measurement bases required by unstructured tomography that is equal to $3^N$. The target states are the same as those in Fig.\,\ref{fig:ca_D}.}
   \label{fig:ca_Nb}
\end{figure}

Fig.\,\ref{fig:ca_Nb} shows the number of measurement bases $N_b$ required by the cross approximation based QST protocol. We obtain $N_b$ in our numerical experiments by counting the number of times a distinct Pauli expectation value $\langle\sigma_{1}^{\gamma_{1}}\sigma_{2}^{\gamma_{2}}\cdots\sigma_{N}^{\gamma_{N}}\rangle$ is used in the DMRG-cross algorithm. As we expect, $N_b$ indeed scales linearly in $N$ (and quadratically in the bond dimension used by DMRG-cross). In contrast, unstructured tomography based on local measurements requires $3^N$ measurement bases, which far exceeds $N_b$ for all $N>8$.

\section{Effects of statistical errors}
The benchmark results in Section II assumes that the elements of the tensor $\mathcal{A}$ are measured exactly. In practice, however, we can only measure $\langle\sigma_{1}^{\gamma_{1}}\sigma_{2}^{\gamma_{2}}\cdots\sigma_{N}^{\gamma_{N}}\rangle$ up a standard deviation of $\delta\le1/\sqrt{M},$where $M$ is
the number of repeated quantum measurements or identical state copies. Note that throughout this work we assume that the measurements are performed on individual copies of the $N$-qubit state and do not consider collective measurements \cite{haah2017sample} that involve measuring multiple state copies simultaneously with possible entangling operations. Collective measurements may require fewer state copies to achieve the same precision in QST \cite{O_Donnell_2016}, but they are very challenging to realize experimentally. 

Let us first estimate the precision $\delta$ needed in measuring $\langle\sigma_{1}^{\gamma_{1}}\sigma_{2}^{\gamma_{2}}\cdots\sigma_{N}^{\gamma_{N}}\rangle$. The basic idea is that $\delta$ should be much smaller than the typical magnitude of $\langle\sigma_{1}^{\gamma_{1}}\sigma_{2}^{\gamma_{2}}\cdots\sigma_{N}^{\gamma_{N}}\rangle$ to ensure a high signal-to-noise ratio. The root mean square value of $\langle\sigma_{1}^{\gamma_{1}}\sigma_{2}^{\gamma_{2}}\cdots\sigma_{N}^{\gamma_{N}}\rangle$ over all values of $\gamma_1,\cdots, \gamma_N$ in a target state $\rho$ can be estimated using:
\begin{equation}
\langle\sigma_{1}^{\gamma_{1}}\cdots\sigma_{N}^{\gamma_{N}}\rangle _{\text{rms}} = \sqrt{\frac{1}{4^N}\sum_{\gamma_{1}\cdots\gamma_{N}}\left|\langle\sigma_{1}^{\gamma_{1}}\cdots\sigma_{N}^{\gamma_{N}}\rangle_{\rho}\right|^{2}} =\sqrt{\frac{\Tr(\rho^2)}{2^N}}.
\end{equation}
Therefore, for the statistical errors to be small, we require
\begin{equation}
\delta = \epsilon \sqrt{\frac{\Tr(\rho^2)}{2^N}} \text{ and } M \ge \frac{1}{\delta^2} = \frac{2^N}{\epsilon^2\Tr(\rho^2)}
\label{eq:M}
\end{equation}
where $\epsilon$ is a small number that quantifies the average relative error in measuring $\langle\sigma_{1}^{\gamma_{1}}\sigma_{2}^{\gamma_{2}}\cdots\sigma_{N}^{\gamma_{N}}\rangle$. Since $2^{-N}\le\Tr(\rho^2)\le1$, the number of state copies per measurement basis scales as $O(2^N)$ to $O(4^N)$, which makes the protocol not scalable. However, we emphasize that this appears to be a fundamental limitation also shared by other QST methods. For example, for an unstructured state with a full rank density operator, the best known protocol for QST using the same local Pauli measurements as ours requires a total number of state copies given by $O(6^N/(\epsilon^2 \Tr(\rho^2)))$ \cite{Gu_2020} for $D=\epsilon^2$ between the target state and the reconstructed state. Thus the number of state copies per basis in such a protocol is the same as our Eq.\,\eqref{eq:M}. Our cross approximation based protocol still has the major advantage of using only $O(N)$ instead of $3^N$ measurement bases, thus requiring a much smaller number of total state copies.

\begin{figure}
   \centering
   \includegraphics[width=0.49\columnwidth]{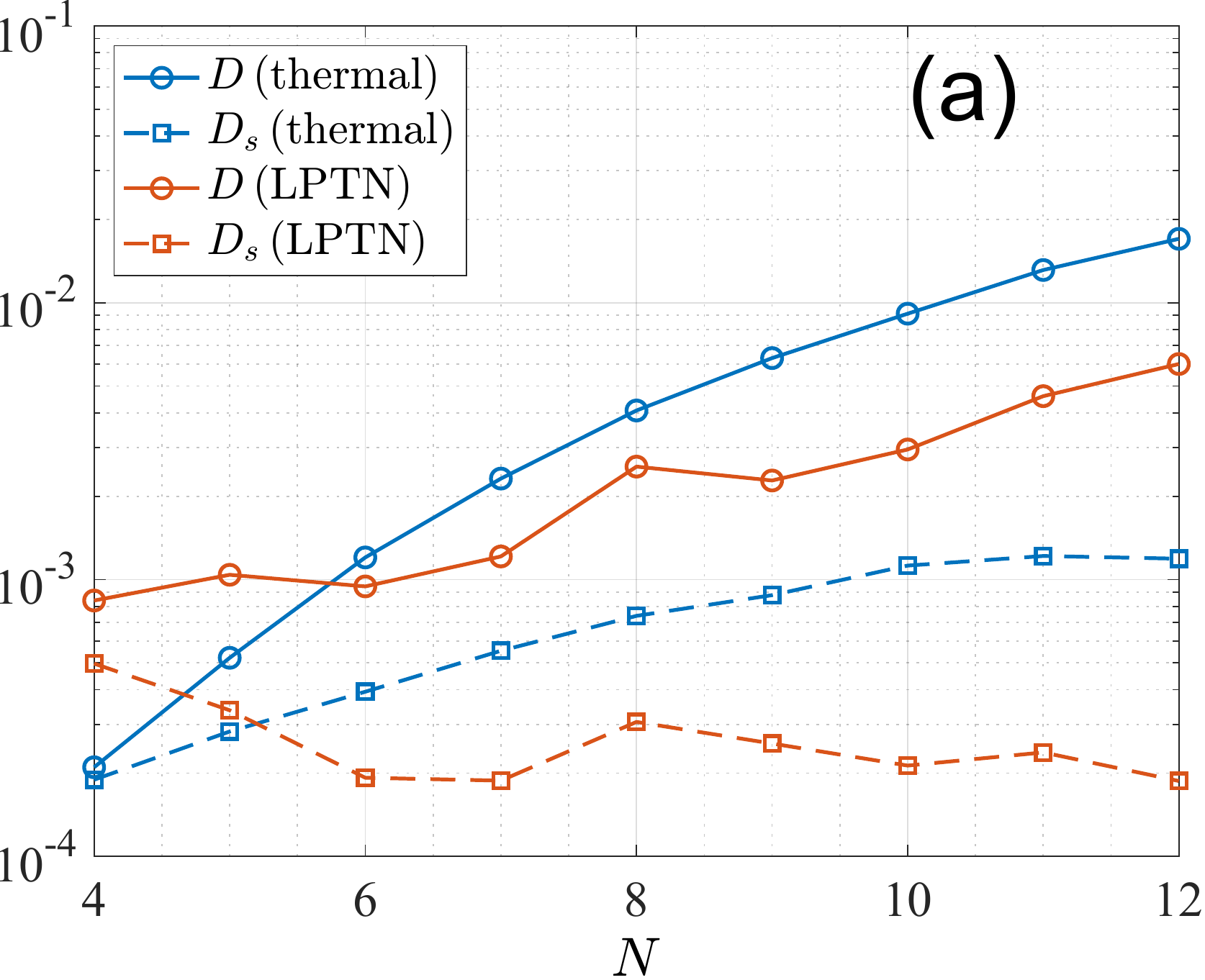} \hfill
   \includegraphics[width=0.49\columnwidth]{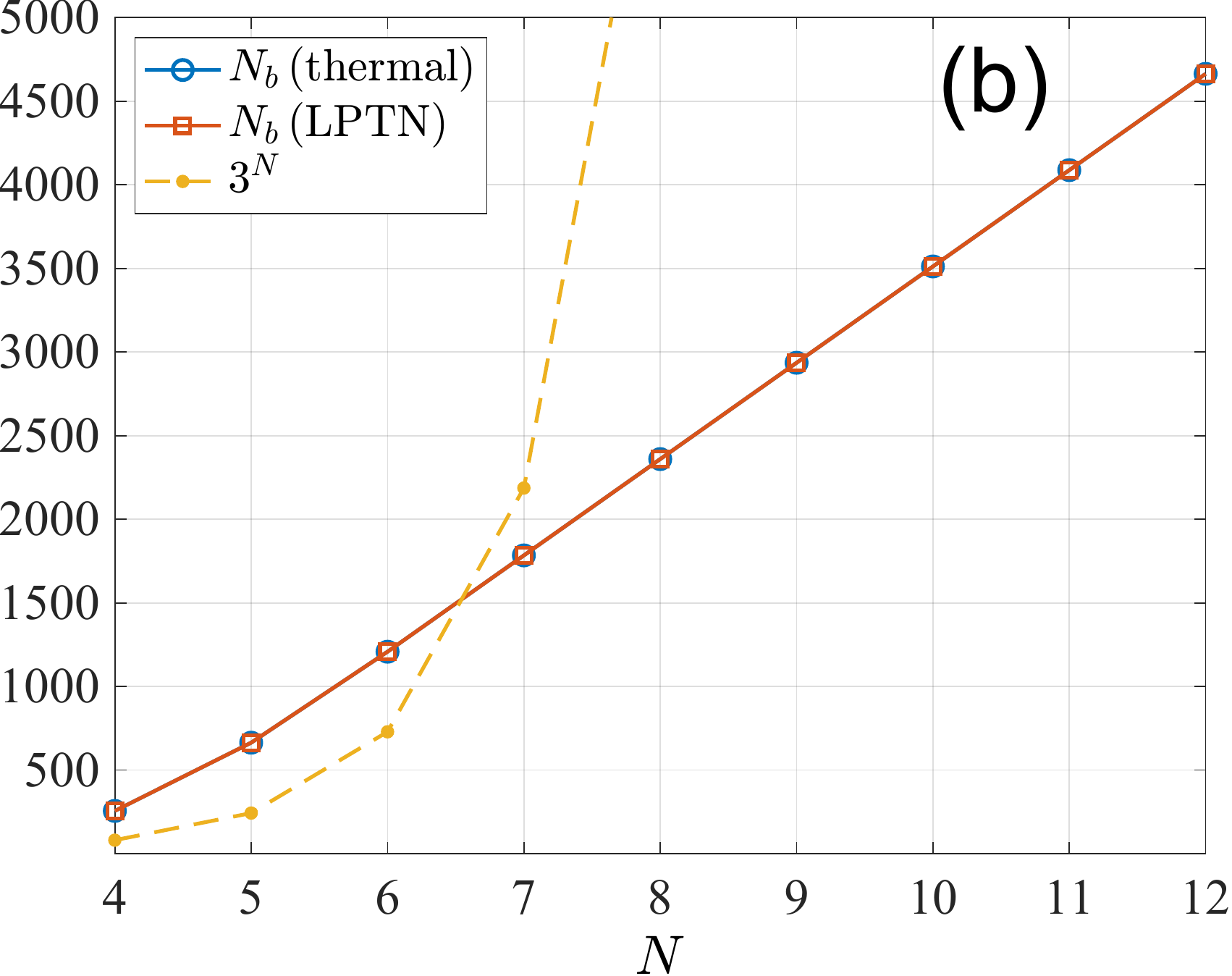}

   \caption{(a) Distance measures ($D$ and $D_s$) between the target states and the reconstructed states using tensor train cross approximation with measurement errors quantified by $\epsilon=0.01$ (see main text). The target states are either thermal states of a quantum Ising model with $T=0.2$ or random LPTN states with $\chi=16$. (b) The number of measurement bases $N_b$ needed for the cross approximation to reconstruct both target states, compared again with $3^N$.}
   \label{fig:ca_err}
\end{figure}

We now show that the DMRG-cross algorithm can indeed tolerate a small amount of measurement error quantified by a small relative error threshold $\epsilon$. Since the number of measurements per basis scales exponentially in $N$, for practical reasons we limit our study to $N\le12$ and the maximum bond dimension in DMRG-cross to $6$. We choose the target states to be either the thermal states of Eq.\,\eqref{eq:HTFIM} with $T=0.2$ or the random LPTN states with bond dimension $\chi=16$, which have been studied in Section II. To simulate the statistical errors on the quantum measurements, we add Gaussian random noise with a standard deviation given by $\delta$ defined in Eq.\,\eqref{eq:M} to each value of $\langle\sigma_{1}^{\gamma_{1}}\sigma_{2}^{\gamma_{2}}\cdots\sigma_{N}^{\gamma_{N}}\rangle$ used in the cross approximation, with $\epsilon=0.01$. Due to such randomness in each numerical experiment, we average $D$ and $N_b$ over 80 repeated experiments. The results are shown in Fig.\,\ref{fig:ca_err}, where we clearly see that the distance measure $D$ is larger than that in Fig.\,\ref{fig:ca_D} without measurement errors. However, the increase of $D$ with $N$ is again slow. This is consistent with our recent work \cite{Zhen_Zhihui_Error} showing that the reconstruction error for tensor train approximation scales at most polynomially in $N$ for a finite relative error $\epsilon$ in the measurements.
 
\section{Improvement using machine learning}
As we mentioned in Section II, each expectation value $\langle\sigma_{1}^{\gamma_{1}}\sigma_{2}^{\gamma_{2}}\cdots\sigma_{N}^{\gamma_{N}}\rangle$ used in the tensor train cross approximation is obtained by measuring $\sigma_i^{\gamma_i}$ in the target state for all qubits. This means each state copy actually provides $N$ bits of information, as each qubit will yield a measurement outcome of either $1$ or $-1$ upon the measurement of $\sigma^{x,y,z}$. However, we only use one bit of information in evaluating $\langle\sigma_{1}^{\gamma_{1}}\sigma_{2}^{\gamma_{2}}\cdots\sigma_{N}^{\gamma_{N}}\rangle$ from each state copy. This means the experimental measurement data contains far more information than what we used in the cross approximation. The full information of the experimental data can instead by captured by the expectation values of all measured operators $\sigma_{1}^{\gamma_{1}}\cdots\sigma_{N}^{\gamma_{N}}$ with no $\gamma_i=0$ plus those with any one or more $\gamma_i$ replaced by zero. Alternatively, we can use the full counting statistics of $\sigma_{1}^{\gamma_{1}},\cdots,\sigma_{N}^{\gamma_{N}}$ to capture the full information of measurement data, which is much more efficient for a large $N$.

\begin{figure}
   \centering
   \includegraphics[width=0.49\columnwidth]{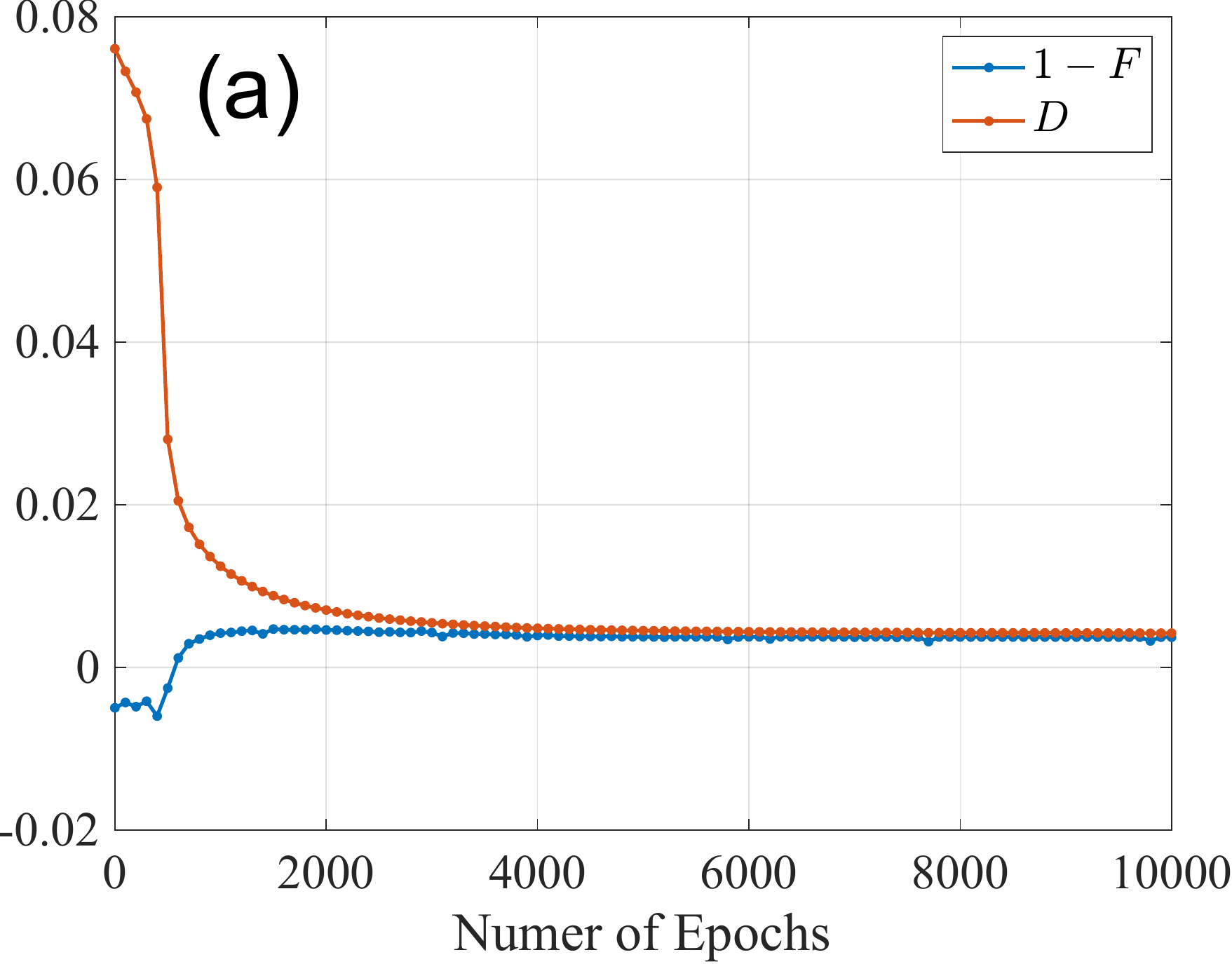} \hfill
   \includegraphics[width=0.49\columnwidth]{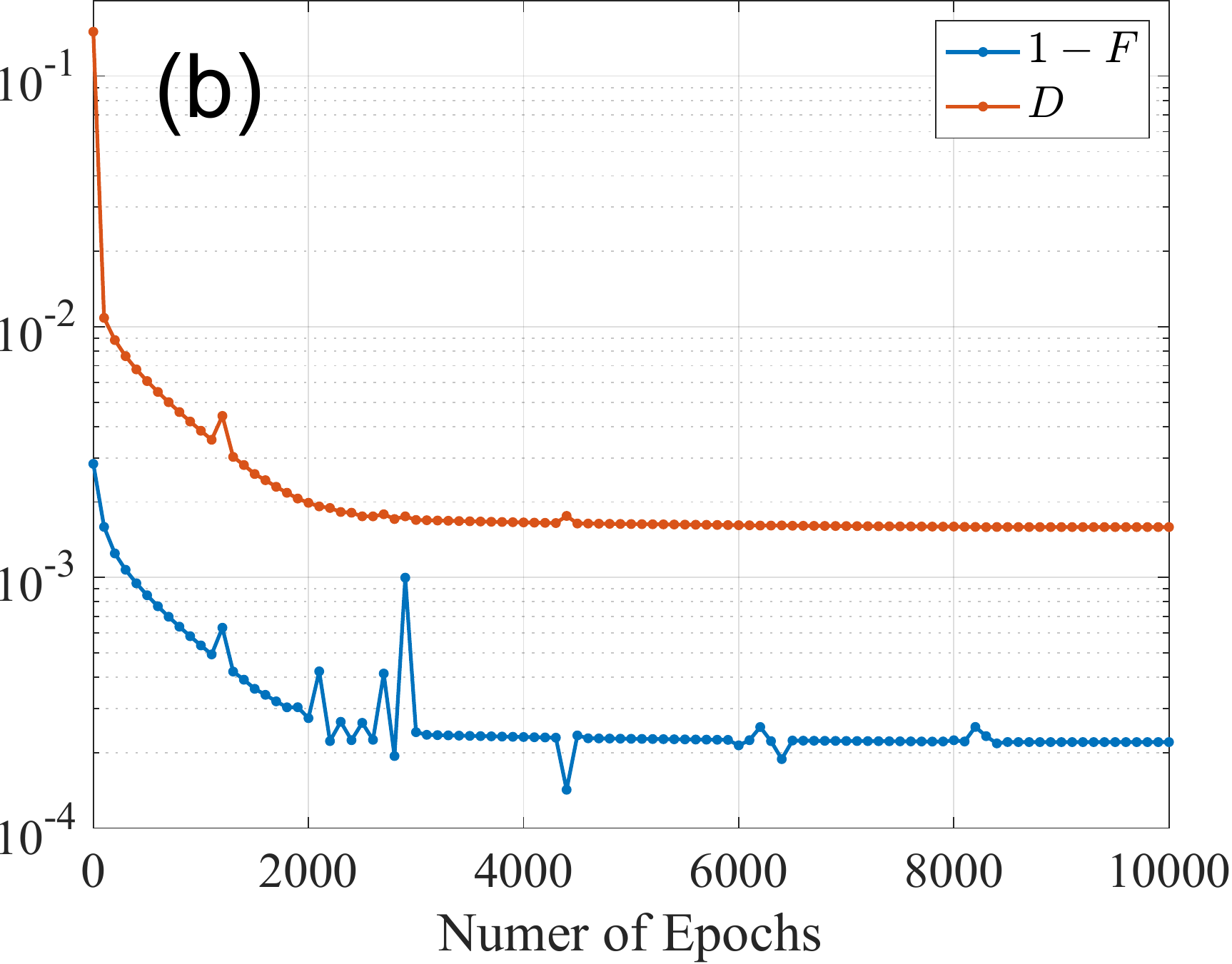}
   \caption{Distance measure $D$ and infidelity $1-F$ between the target states and the reconstructed states obtained using tensor train cross approximation and further trained via supervised machine learning. Statistical errors are included by simulating $M=10^6$ repeated measurements per basis. The target state is either a thermal state of a quantum Ising model with $T=1$ (a) or a random LPTN state with $\chi=16$ (b), both with $N=8$ qubits. The horizontal axes represent the training process.}
   \label{fig:ml}
\end{figure}

In this section, we show that such extra information contained in the experimental data but unused by the cross approximation can be harnessed to improve the quality of the QST via a supervised machine learning procedure. Our idea is simple: We can use the MPO $\{\tilde{G}_i^{\gamma_i}\}$ obtained using the tensor train cross approximation as a good initial guess, and use the extra entries of $\mathcal{A}$ calculated from the same experimental data to refine $\{\tilde{G}_i^{\gamma_i}\}$ via supervised learning. The loss function during the training is defined as
\begin{equation}
    L = \sideset{}{''}\sum_{\gamma_1,\cdots,\gamma_N}|\langle\sigma_{1}^{\gamma_{1}}\cdots\sigma_{N}^{\gamma_{N}} \rangle_{\rho_1}-\langle\sigma_{1}^{\gamma_{1}}\cdots\sigma_{N}^{\gamma_{N}} \rangle_{\rho_2}|^2
\end{equation}
where $\sideset{}{''}\sum_{\gamma_1,\cdots,\gamma_N}$ denotes the summation over all indices $\gamma_1,\cdots,\gamma_N$ of which $\langle\sigma_{1}^{\gamma_{1}}\sigma_{2}^{\gamma_{2}}\cdots\sigma_{N}^{\gamma_{N}}\rangle$ can be computed from the experimental data, each subject to statistical errors quantified in Section III. $\rho_1$ here denotes the experimental state being measured and $\rho_2$ denotes the state represented by the MPO being trained. Importantly, $L$ can be calculated just based on the experimental data, without knowing the full target state. We then perform a stochastic gradient descent on the loss function $L$ over the parameters in the MPO being trained to minimize $L$ using an adaptive moment estimation (Adam) method \cite{ADAM}. 

To benchmark this supervised machine learning method, we set $N=8$ and use either a target state that is a thermal state of Eq.\,\eqref{eq:HTFIM} with $T=1$ or a random LPTN state with bond dimension $\chi=16$. We use $M=10^6$ repeated measurements per basis and a total number of roughly $1000$ measurement bases as required by the cross approximation. For both target states, the tensor train cross approximation gives us a reconstructed MPO with $D\approx 0.1$, which is reasonably good but not ideal due to a finite statistic error. As shown in Fig.\,\ref{fig:ml}, supervised machine learning is able to refine the MPO obtained using cross approximation significantly. For the thermal state, the distance measure $D$ drops by a factor of about $20$, while for the random LPTN state $D$ drops by a factor of almost $100$. Since the system size is small, here we can also compute the infidelity $1-F$ between the target state $\rho_1$ and the reconstructed state $\rho_2$ (obtained from their MPO representations), with the fidelity defined as $F\equiv \Tr\sqrt{\sqrt{\rho_1} \rho_2 \sqrt{\rho_1}}$. For the thermal state, the infidelity is negative initially, showing that the reconstructed density matrix is not strictly semi-positive definite. But the supervised machine learning is able to correct this and produces a positive density matrix with a fidelity of $F\approx 99.6\%$ at the end of the training. For the random LPTN state, we were able to improve the fidelity from $99.7\%$ initially to $99.98\%$.

\section{Discussion and Outlook}
The supervised machine learning introduced in the previous section can noticeably reduce the number of repeated measurements $M$ per basis while maintaining the same level of reconstruction error. However, we do not expect it to overcome the exponential scaling of $M$ in the system size $N$. To make the QST fully scalable, we can combine the tensor train cross approximation method with the MPO concatenation method in Refs.\,\cite{baumgratz2013scalable,Baumgratz_2013}. First, we note that if an $N$-qubit mixed state can be represented by an MPO with a maximum bond dimension $\chi$, then any $R$-qubit reduced state can also be represented by an MPO with a bond dimension of at most $\chi$. As a result, if the MPO for the full state satisfies the invertibility condition in Ref.\,\cite{baumgratz2013scalable} such that it can be reconstructed via the reduced states of $R$ consecutive qubits, we can use the tensor train cross approximation to perform QST on $O(N)$ of such $R$ consecutive qubits, as long as $R$ is not too large. This is a significant improvement over tomography methods that do not take advantage of the structure of the reduced states and hence require exponentially more (in $N$) measurements. Such improvement can be seen for $R$ as small as $7$ based on the examples we studied (see Fig.\,\ref{fig:ca_err}b).

We emphasize that to our best knowledge, no provably efficient QST protocol has been found for generic physical states with a compact MPO representation. Our work highlights an important roadblock towards finding such efficient protocol. Although a compact MPO state has only $O(N)$ independent parameters, the state in general contains an exponentially large (in $N$) number of nonzero parameters and thus each parameter has an exponentially small typical value. Measuring even $O(N)$ such parameters to a small relative error would thus require an exponentially large number of measurements. This fundamental limitation seems only avoidable if the full state can be obtained by local reductions such that one just need to perform QST on small subsystems instead. However, the existence of such local reductions cannot be guaranteed \cite{baumgratz2013scalable,Baumgratz_2013}. A very useful future direction is to derive upper bounds on the size $R$ in the local reductions, provided that the target state is represented by an MPO with a finite bond dimension. This may be possible especially if we can tolerate a small amount of error in the reconstruction of the target state. 

On the other hand, one can try QST protocols where no single parameter of the density matrix is measured to a good precision. For example, Ref.\,\cite{wang2020scalable} shows that one can perform just one measurement per basis and choose a large number of random local bases to reconstruct an MPS target state very efficiently. This is an orthogonal approach from ours that requires a large number of measurements per basis but a minimal number of bases. However, how to generalize this approach to mixed states represented by MPOs is another interesting open question. In addition, instead of choosing the measurement bases randomly in such an approach, one can adaptively choose new bases based on the previous measurement outcomes. Such adaptive measurement protocols have been shown to perform better in QST \cite{haah2017sample,lange2022adaptive} than those using independent measurement bases. In fact, the maximum volume principle based tensor cross approximation used in this work is already such an adaptive protocol. Whether adaptive measurement protocols can eventually lead to efficient QST for generic MPO target states remains to be seen.

\begin{acknowledgments}
We thank the HPC center at Colorado School of Mines for providing computational resources needed in carrying out this work. We acknowledge funding support from NSF Grants No. CCF-1839232, PHY-2112893, CCF-2106834 and CCF-2106881, as well as the W. M. Keck Foundation.
\end{acknowledgments}

\bibliographystyle{apsrev4-2}
\bibliography{refs}

\end{document}